\documentclass[sigconf]{acmart}
\AtBeginDocument{%
  \providecommand\BibTeX{{%
    \normalfont B\kern-0.5em{\scshape i\kern-0.25em b}\kern-0.8em\TeX}}}

\setcopyright{rightsretained}

\acmConference[SIGCHI 2023]{The Second Workshop on Intelligent and Interactive Writing Assistants}{April 23--28,
  2023}{Hamburg, Germany}
\setcopyright{none}
\renewcommand\footnotetextcopyrightpermission[1]{}
\begin{document}
\title{Approach Intelligent Writing Assistants Usability with Seven Stages of Action}

\author{Avinash Bhat}
\affiliation{\institution{McGill University} \country{Canada}}
\author{Disha Shrivastava}
\affiliation{\institution{Mila, Université de Montréal} \country{Canada}}
\author{Jin L.C. Guo}
\affiliation{\institution{McGill University} \country{Canada}}

\begin{abstract}
Despite the potential of Large Language Models (LLMs) as writing assistants, they are plagued by issues like coherence and fluency of the model output, trustworthiness, ownership of the generated content, and predictability of model performance, thereby limiting their usability. In this position paper, we propose to adopt Norman's seven stages of action as a framework to approach the interaction design of intelligent writing assistants. We illustrate the framework's applicability to writing tasks by providing an example of software tutorial authoring. The paper also discusses the framework as a tool to synthesize research on the interaction design of LLM-based tools and presents examples of tools that support the stages of action. Finally, we briefly outline the potential of a framework for human-LLM interaction research.
\end{abstract}

\begin{teaserfigure}
  \includegraphics[width=\textwidth]{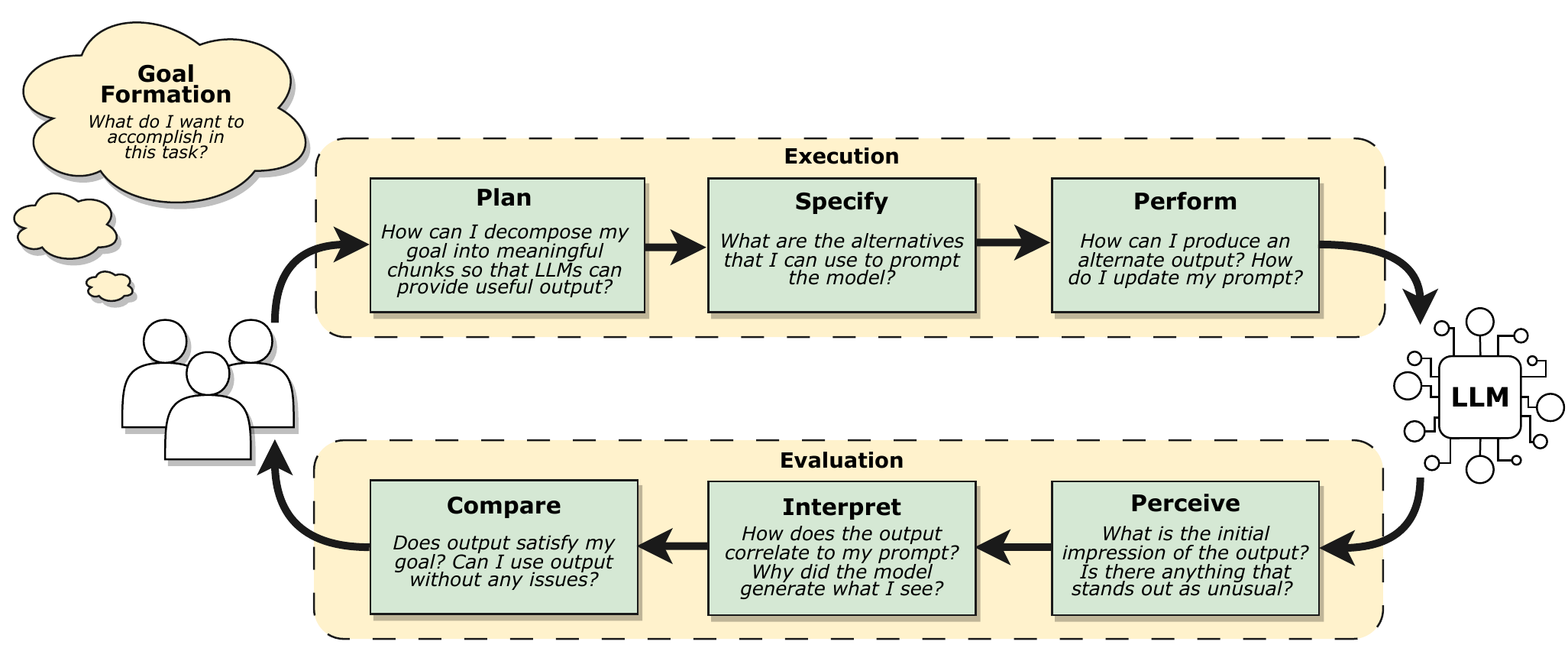}
  \caption{Depiction of Norman's Seven Stages of Action examined in the context of interactions with the LLM.}
  \label{fig1}
\end{teaserfigure}

\maketitle
\section{Introduction}
Intelligent writing assistants have been widely explored for various writing goals and activities~\cite{Gero2022Design}. The recent progress in writing assistants has been centred around Large Language Models (LLMs)~\cite{Coenen2021Wordcraft}, using which humans can generate content following the intent provided as a prompt. The notable advancements in LLMs like ChatGPT\footnote{https://openai.com/blog/chatgpt} and its adoption in everyday products\footnote{https://openai.com/blog/chatgpt-plugins} highlight their potential as writing assistants. However, the human interaction with such assistants exposes major limitations related to their usability, such as coherence and fluency of the model output~\cite{Yang2022ActiveWriter, Ghajargar2022Redhead}, trustworthiness~\cite{Goodman2022Lampost}, as well as ownership~\cite{Biermann2022Companion} of the generated content and the predictability of model performance~\cite{Ippolito2022CreativeWW, Ghajargar2022Redhead}. These issues often result in users being unable to use the tools effectively to achieve their writing goals and sometimes abandoning them entirely. 

While previous works~\cite{Gero2022Design, Chung2021Intersection, Lee2022Evaluating} have investigated the interaction aspects of writing assistants to some extent, there is no dedicated effort in meeting end-to-end writing objectives and approaching their interactions from a usability standpoint. We draw inspiration from these studies and existing design literature to investigate the interaction design in intelligent writing assistants supported by LLMs with a focus on human actions. We further propose to adopt Norman's seven stages of action~\cite{Norman2002DOET} as a framework to guide the design of LLM-supported intelligent writing assistants and discuss its implication on usability.

\section{Seven Stages of Action}
Norman’s seven stages of action is a cyclical cognitive model commonly used to comprehend the users' mental processes and corresponding physical action primarily employed to guide the interaction design of a system. As depicted in Figure \ref{fig1}, the seven stages of action consists of (a) \emph{goal formation}, (b) \emph{plan}, (c) \emph{specify}, (d) \emph{perform}, (e) \emph{perceive}, (f) \emph{interpret}, and (g) \emph{compare}. The plan, specify, and perform stages form the \emph{execution phase}, and perceive, interpret and compare stages form the \emph{evaluation phase} of the interaction. The user's interactions depend on a \emph{mental model} of the system formed by users' prior beliefs. We posit that this framework offers the opportunity to design and evaluate interfaces that support fine-grained actions at different stages when interacting with LLMs. We propose that an effective design for LLM-based writing assistants must answer the questions pertaining to the different stages in order to guide the design and provide the necessary capabilities to the user. 

To illustrate our idea further, we provide an example largely inspired by our initial attempt to leverage OpenAI's Codex~\cite{Chen2021Codex} in the task of writing software tutorials. In a typical interaction, the user starts by forming an initial \textit{goal} that they want to accomplish, e.g., to write a tutorial for plotting data points with matplotlib. Next, they \textit{plan} the goal by dividing it into pertinent parts which can guide them towards querying the writing assistant. For example, the overall goal can be divided into first writing tutorial sections such as producing relevant commands for library installation in different environments, then generating and explaining code snippets, and lastly improving the readability of the tutorial. Here, an individual step can also be considered a \textit{sub-goal}, albeit with a smaller scope, and can follow multiple iterations of the action framework. When the users prompt the writing assistant, they tend to \textit{specify} and later \textit{perform} their requirement in the writing assistant interface, e.g., 
“Write a code snippet to plot a scatter plot using matplotlib given the data points in a Python list and provide an explanation of the code.”. The \emph{specify} stage can have mechanisms to suggest alternate prompts to the model and the \emph{perform} stage can have different interface features to edit and update the prompts. The users' knowledge about the task and domain, and their existing conceptual models inform the execution phase. Once the writing assistant returns an output, the user \textit{perceives} and \textit{interprets} it according to their knowledge and expertise and updates their existing mental models. For example, a user with extensive knowledge of using matplotlib might be able to better perceive any unusual content or errors in the generated code. It might also be necessary to \textit{compare} the output with resources in different environments, e.g., by executing the generated code snippet in an IDE or running any existing unit tests. Asking questions that are relevant to each stage can identify the crucial interactions and guide the design of a writing assistant to support the tutorial authoring task.

\section{Discussion}
Norman's seven stages of action framework complement the Cognitive Process Theory of Writing~\cite{Flower1981CPT} as discussed by~\citet{Gero2022Design} in their study on the design space of writing assistants. The Cognitive Process Theory of Writing identifies several key processes that occur during the writing process such as generating ideas, organizing and setting writing goals, and revising the written content. The stages of action can be introduced during each of these cognitive processes to incorporate LLMs as complementary agents that assist the writers. Given the effectiveness exhibited by the LLMs in different writing tasks, it might be necessary to consider interaction as a dimension while studying the design space of writing tools. Notably, a recent study by \citet{Lee2022Evaluating} evaluated the LLMs based on their performance in a human-LLM interactive setting and emphasized the importance of investigating such interactions to better assess the model. Inspired by these studies, we propose introducing the seven stages of action framework to extend the discussion around capabilities and interactions required in LLM-based tools and the corresponding implications.

The seven stages of action framework facilitates the synthesis of existing research on the interaction strategies employed in LLM-based tools, indicating a (perhaps undesirable) emphasis on the execution stage. Existing strategies such as chaining prompts \cite{Wu2022AIChains}, which addresses how to design prompts, can be considered as the effort supporting the planning and specification stages. Another example of a tool supporting the execution phase is PromptMaker \cite{Jiang2022PromptMaker} which provides alternative interfaces for designing prompts. 

Specifically, in the case of interactions in LLM-based writing assistants, the meta-prompting strategy in Wordcraft \cite{Coenen2021Wordcraft} provides alternative prompts to the users, assisting in the specification stage. TaleBrush \cite{Chung2022TaleBrush} introduces an alternative sketch input to the model for steering the content generation, demonstrating the interaction possibilities in the specify and perform stages. In the evaluation phase, PromptChainer \cite{Wu2022PromptChainer} visualizes the output at every step in the prompt design, enabling better interaction cycles by providing effective mappings between the user specification and LLM output. 

Viewing the human-LLM interaction through the lens of the seven stages of action framework can potentially help to better manage and address the associated challenges. For instance, using LLMs as a backend for writing assistants comes with risks~\cite{Weidinger2022Taxonomy} involving fairness issues like discrimination based on social stereotypes and targeted hate speech. While there are strategies to identify~\cite{Markov2023Holistic} and limit\footnote{https://platform.openai.com/docs/guides/moderation} certain cases of misuse, it would be extremely challenging, if not impossible, to entirely moderate the interaction due to its open-ended nature~\cite{Ganguli2022Predictability}. In some situations, moderation might even be unacceptable considering the fine line between creative and objectionable output~\cite{Ippolito2022CreativeWW}. Understanding current mental models of human-LLM interaction and decomposing them into stages with specific motives can potentially facilitate content moderation. For example, an assistant that employs the seven stages can leverage the \emph{plan} stage to define and moderate the scope of the output and the \emph{specify} stage to supervise the prompt and ensure the user does not request any undesirable content. \emph{Interpret} and \emph{compare} stages can be used to verify the relevance of the output and mitigate any undesired content.

One of the challenges that we foresee with using the framework is that the distinction between the seven stages is not always apparent, especially when dealing with complex tasks such as writing. For instance, it can be difficult to discern the boundary between the \emph{perceive} and \emph{interpret} stages of the output, as these processes often occur simultaneously and interactively. A rigid adherence to the framework may not be feasible in all cases. To address this limitation, the designers can adopt a flexible approach that takes the unique characteristics of the task and the user's needs into account and construe the designs specifically to the task. Depending on the use case, the designers can identify certain stages to pay more attention to or realize the stages differently.

For example, while the compare stage for assistants dealing with software tutorials and research writing both might be trying to address the same design goal of ensuring the output is accurate, the way to address it for the former use case can be by executing the code/tutorial while the latter use case can address it by providing links to digital libraries for verifiable information. Any nuances and task-specific requirements can potentially be identified by using user-centric design methods, such as user interviews or participatory design and can be leveraged by the designers to create solutions that are tailored to the goal.

\textbf{Conclusion.} We maintain that the adoption of Norman's seven stages of action as a framework to explore user actions with LLM-based writing assistants can provide a valuable structure to realize and design fine-grained interactions across goal formation, execution, and  evaluation phases. Analyzing the tools and their features across the interaction design dimensions laid out by the framework can address specific usability concerns in the design of LLM-based writing tools. More ambitiously, they indicate potential avenues for research in human-LLM interactions that are currently underrepresented, such as alignment to human preferences~\cite{Christiano2017RLHF}, effective prompt design~\cite{Liu2023PPP} and explainability and interpretability of model outputs \cite{Guidotti2018XAISurvey}. 

\bibliographystyle{ACM-Reference-Format} 
\bibliography{references}


\begin{thebibliography}{22}


\ifx \showCODEN    \undefined \def \showCODEN     #1{\unskip}     \fi
\ifx \showDOI      \undefined \def \showDOI       #1{#1}\fi
\ifx \showISBNx    \undefined \def \showISBNx     #1{\unskip}     \fi
\ifx \showISBNxiii \undefined \def \showISBNxiii  #1{\unskip}     \fi
\ifx \showISSN     \undefined \def \showISSN      #1{\unskip}     \fi
\ifx \showLCCN     \undefined \def \showLCCN      #1{\unskip}     \fi
\ifx \shownote     \undefined \def \shownote      #1{#1}          \fi
\ifx \showarticletitle \undefined \def \showarticletitle #1{#1}   \fi
\ifx \showURL      \undefined \def \showURL       {\relax}        \fi
\providecommand\bibfield[2]{#2}
\providecommand\bibinfo[2]{#2}
\providecommand\natexlab[1]{#1}
\providecommand\showeprint[2][]{arXiv:#2}

\bibitem[Biermann et~al\mbox{.}(2022)]%
        {Biermann2022Companion}
\bibfield{author}{\bibinfo{person}{Oloff~C. Biermann}, \bibinfo{person}{Ning~F.
  Ma}, {and} \bibinfo{person}{Dongwook Yoon}.} \bibinfo{year}{2022}\natexlab{}.
\newblock \showarticletitle{From Tool to Companion: Storywriters Want AI
  Writers to Respect Their Personal Values and Writing Strategies}. In
  \bibinfo{booktitle}{\emph{Designing Interactive Systems Conference}} (Virtual
  Event, Australia) \emph{(\bibinfo{series}{DIS '22})}.
  \bibinfo{publisher}{Association for Computing Machinery},
  \bibinfo{address}{New York, NY, USA}, \bibinfo{pages}{1209–1227}.
\newblock
\showISBNx{9781450393584}
\urldef\tempurl%
\url{https://doi.org/10.1145/3532106.3533506}
\showDOI{\tempurl}


\bibitem[Chen et~al\mbox{.}(2021)]%
        {Chen2021Codex}
\bibfield{author}{\bibinfo{person}{Mark Chen}, \bibinfo{person}{Jerry Tworek},
  \bibinfo{person}{Heewoo Jun}, \bibinfo{person}{Qiming Yuan},
  \bibinfo{person}{Henrique Ponde de~Oliveira Pinto}, \bibinfo{person}{Jared
  Kaplan}, \bibinfo{person}{Harri Edwards}, \bibinfo{person}{Yuri Burda},
  \bibinfo{person}{Nicholas Joseph}, \bibinfo{person}{Greg Brockman},
  \bibinfo{person}{Alex Ray}, \bibinfo{person}{Raul Puri},
  \bibinfo{person}{Gretchen Krueger}, \bibinfo{person}{Michael Petrov},
  \bibinfo{person}{Heidy Khlaaf}, \bibinfo{person}{Girish Sastry},
  \bibinfo{person}{Pamela Mishkin}, \bibinfo{person}{Brooke Chan},
  \bibinfo{person}{Scott Gray}, \bibinfo{person}{Nick Ryder},
  \bibinfo{person}{Mikhail Pavlov}, \bibinfo{person}{Alethea Power},
  \bibinfo{person}{Lukasz Kaiser}, \bibinfo{person}{Mohammad Bavarian},
  \bibinfo{person}{Clemens Winter}, \bibinfo{person}{Philippe Tillet},
  \bibinfo{person}{Felipe~Petroski Such}, \bibinfo{person}{Dave Cummings},
  \bibinfo{person}{Matthias Plappert}, \bibinfo{person}{Fotios Chantzis},
  \bibinfo{person}{Elizabeth Barnes}, \bibinfo{person}{Ariel Herbert-Voss},
  \bibinfo{person}{William~Hebgen Guss}, \bibinfo{person}{Alex Nichol},
  \bibinfo{person}{Alex Paino}, \bibinfo{person}{Nikolas Tezak},
  \bibinfo{person}{Jie Tang}, \bibinfo{person}{Igor Babuschkin},
  \bibinfo{person}{Suchir Balaji}, \bibinfo{person}{Shantanu Jain},
  \bibinfo{person}{William Saunders}, \bibinfo{person}{Christopher Hesse},
  \bibinfo{person}{Andrew~N. Carr}, \bibinfo{person}{Jan Leike},
  \bibinfo{person}{Josh Achiam}, \bibinfo{person}{Vedant Misra},
  \bibinfo{person}{Evan Morikawa}, \bibinfo{person}{Alec Radford},
  \bibinfo{person}{Matthew Knight}, \bibinfo{person}{Miles Brundage},
  \bibinfo{person}{Mira Murati}, \bibinfo{person}{Katie Mayer},
  \bibinfo{person}{Peter Welinder}, \bibinfo{person}{Bob McGrew},
  \bibinfo{person}{Dario Amodei}, \bibinfo{person}{Sam McCandlish},
  \bibinfo{person}{Ilya Sutskever}, {and} \bibinfo{person}{Wojciech Zaremba}.}
  \bibinfo{year}{2021}\natexlab{}.
\newblock \bibinfo{title}{Evaluating Large Language Models Trained on Code}.
\newblock
\newblock
\urldef\tempurl%
\url{https://doi.org/10.48550/ARXIV.2107.03374}
\showDOI{\tempurl}


\bibitem[Christiano et~al\mbox{.}(2017)]%
        {Christiano2017RLHF}
\bibfield{author}{\bibinfo{person}{Paul Christiano}, \bibinfo{person}{Jan
  Leike}, \bibinfo{person}{Tom~B. Brown}, \bibinfo{person}{Miljan Martic},
  \bibinfo{person}{Shane Legg}, {and} \bibinfo{person}{Dario Amodei}.}
  \bibinfo{year}{2017}\natexlab{}.
\newblock \bibinfo{title}{Deep reinforcement learning from human preferences}.
\newblock
\newblock
\urldef\tempurl%
\url{https://doi.org/10.48550/ARXIV.1706.03741}
\showDOI{\tempurl}


\bibitem[Chung et~al\mbox{.}(2021)]%
        {Chung2021Intersection}
\bibfield{author}{\bibinfo{person}{John Joon~Young Chung},
  \bibinfo{person}{Shiqing He}, {and} \bibinfo{person}{Eytan Adar}.}
  \bibinfo{year}{2021}\natexlab{}.
\newblock \showarticletitle{The Intersection of Users, Roles, Interactions, and
  Technologies in Creativity Support Tools}. In
  \bibinfo{booktitle}{\emph{Designing Interactive Systems Conference 2021}}
  (Virtual Event, USA) \emph{(\bibinfo{series}{DIS '21})}.
  \bibinfo{publisher}{Association for Computing Machinery},
  \bibinfo{address}{New York, NY, USA}, \bibinfo{pages}{1817–1833}.
\newblock
\showISBNx{9781450384766}
\urldef\tempurl%
\url{https://doi.org/10.1145/3461778.3462050}
\showDOI{\tempurl}


\bibitem[Chung et~al\mbox{.}(2022)]%
        {Chung2022TaleBrush}
\bibfield{author}{\bibinfo{person}{John Joon~Young Chung},
  \bibinfo{person}{Wooseok Kim}, \bibinfo{person}{Kang~Min Yoo},
  \bibinfo{person}{Hwaran Lee}, \bibinfo{person}{Eytan Adar}, {and}
  \bibinfo{person}{Minsuk Chang}.} \bibinfo{year}{2022}\natexlab{}.
\newblock \showarticletitle{TaleBrush: Sketching Stories with Generative
  Pretrained Language Models}. In \bibinfo{booktitle}{\emph{Proceedings of the
  2022 CHI Conference on Human Factors in Computing Systems}} (New Orleans, LA,
  USA) \emph{(\bibinfo{series}{CHI '22})}. \bibinfo{publisher}{Association for
  Computing Machinery}, \bibinfo{address}{New York, NY, USA}, Article
  \bibinfo{articleno}{209}, \bibinfo{numpages}{19}~pages.
\newblock
\showISBNx{9781450391573}
\urldef\tempurl%
\url{https://doi.org/10.1145/3491102.3501819}
\showDOI{\tempurl}


\bibitem[Flower and Hayes(1981)]%
        {Flower1981CPT}
\bibfield{author}{\bibinfo{person}{Linda Flower} {and} \bibinfo{person}{John~R.
  Hayes}.} \bibinfo{year}{1981}\natexlab{}.
\newblock \showarticletitle{A Cognitive Process Theory of Writing}.
\newblock \bibinfo{journal}{\emph{College Composition and Communication}}
  \bibinfo{volume}{32}, \bibinfo{number}{4} (\bibinfo{year}{1981}),
  \bibinfo{pages}{365--387}.
\newblock
\showISSN{0010096X}
\urldef\tempurl%
\url{http://www.jstor.org/stable/356600}
\showURL{%
\tempurl}


\bibitem[Ganguli et~al\mbox{.}(2022)]%
        {Ganguli2022Predictability}
\bibfield{author}{\bibinfo{person}{Deep Ganguli}, \bibinfo{person}{Danny
  Hernandez}, \bibinfo{person}{Liane Lovitt}, \bibinfo{person}{Amanda Askell},
  \bibinfo{person}{Yuntao Bai}, \bibinfo{person}{Anna Chen},
  \bibinfo{person}{Tom Conerly}, \bibinfo{person}{Nova Dassarma},
  \bibinfo{person}{Dawn Drain}, \bibinfo{person}{Nelson Elhage},
  \bibinfo{person}{Sheer El~Showk}, \bibinfo{person}{Stanislav Fort},
  \bibinfo{person}{Zac Hatfield-Dodds}, \bibinfo{person}{Tom Henighan},
  \bibinfo{person}{Scott Johnston}, \bibinfo{person}{Andy Jones},
  \bibinfo{person}{Nicholas Joseph}, \bibinfo{person}{Jackson Kernian},
  \bibinfo{person}{Shauna Kravec}, \bibinfo{person}{Ben Mann},
  \bibinfo{person}{Neel Nanda}, \bibinfo{person}{Kamal Ndousse},
  \bibinfo{person}{Catherine Olsson}, \bibinfo{person}{Daniela Amodei},
  \bibinfo{person}{Tom Brown}, \bibinfo{person}{Jared Kaplan},
  \bibinfo{person}{Sam McCandlish}, \bibinfo{person}{Christopher Olah},
  \bibinfo{person}{Dario Amodei}, {and} \bibinfo{person}{Jack Clark}.}
  \bibinfo{year}{2022}\natexlab{}.
\newblock \showarticletitle{Predictability and Surprise in Large Generative
  Models}. In \bibinfo{booktitle}{\emph{2022 ACM Conference on Fairness,
  Accountability, and Transparency}} (Seoul, Republic of Korea)
  \emph{(\bibinfo{series}{FAccT '22})}. \bibinfo{publisher}{Association for
  Computing Machinery}, \bibinfo{address}{New York, NY, USA},
  \bibinfo{pages}{1747–1764}.
\newblock
\showISBNx{9781450393522}
\urldef\tempurl%
\url{https://doi.org/10.1145/3531146.3533229}
\showDOI{\tempurl}


\bibitem[Gero et~al\mbox{.}(2022)]%
        {Gero2022Design}
\bibfield{author}{\bibinfo{person}{Katy Gero}, \bibinfo{person}{Alex
  Calderwood}, \bibinfo{person}{Charlotte Li}, {and} \bibinfo{person}{Lydia
  Chilton}.} \bibinfo{year}{2022}\natexlab{}.
\newblock \showarticletitle{A Design Space for Writing Support Tools Using a
  Cognitive Process Model of Writing}. In \bibinfo{booktitle}{\emph{Proceedings
  of the First Workshop on Intelligent and Interactive Writing Assistants
  (In2Writing 2022)}}. \bibinfo{publisher}{Association for Computational
  Linguistics}, \bibinfo{address}{Dublin, Ireland}, \bibinfo{pages}{11--24}.
\newblock
\urldef\tempurl%
\url{https://doi.org/10.18653/v1/2022.in2writing-1.2}
\showDOI{\tempurl}


\bibitem[Ghajargar et~al\mbox{.}(2022)]%
        {Ghajargar2022Redhead}
\bibfield{author}{\bibinfo{person}{Maliheh Ghajargar}, \bibinfo{person}{Jeffrey
  Bardzell}, {and} \bibinfo{person}{Love Lagerkvist}.}
  \bibinfo{year}{2022}\natexlab{}.
\newblock \showarticletitle{A Redhead Walks into a Bar: Experiences of Writing
  Fiction with Artificial Intelligence}. In
  \bibinfo{booktitle}{\emph{Proceedings of the 25th International Academic
  Mindtrek Conference}} (Tampere, Finland) \emph{(\bibinfo{series}{Academic
  Mindtrek '22})}. \bibinfo{publisher}{Association for Computing Machinery},
  \bibinfo{address}{New York, NY, USA}, \bibinfo{pages}{230–241}.
\newblock
\showISBNx{9781450399555}
\urldef\tempurl%
\url{https://doi.org/10.1145/3569219.3569418}
\showDOI{\tempurl}


\bibitem[Goodman et~al\mbox{.}(2022)]%
        {Goodman2022Lampost}
\bibfield{author}{\bibinfo{person}{Steven~M. Goodman}, \bibinfo{person}{Erin
  Buehler}, \bibinfo{person}{Patrick Clary}, \bibinfo{person}{Andy Coenen},
  \bibinfo{person}{Aaron Donsbach}, \bibinfo{person}{Tiffanie~N. Horne},
  \bibinfo{person}{Michal Lahav}, \bibinfo{person}{Robert MacDonald},
  \bibinfo{person}{Rain~Breaw Michaels}, \bibinfo{person}{Ajit Narayanan},
  \bibinfo{person}{Mahima Pushkarna}, \bibinfo{person}{Joel Riley},
  \bibinfo{person}{Alex Santana}, \bibinfo{person}{Lei Shi},
  \bibinfo{person}{Rachel Sweeney}, \bibinfo{person}{Phil Weaver},
  \bibinfo{person}{Ann Yuan}, {and} \bibinfo{person}{Meredith~Ringel Morris}.}
  \bibinfo{year}{2022}\natexlab{}.
\newblock \showarticletitle{LaMPost: Design and Evaluation of an AI-Assisted
  Email Writing Prototype for Adults with Dyslexia}. In
  \bibinfo{booktitle}{\emph{Proceedings of the 24th International ACM SIGACCESS
  Conference on Computers and Accessibility}} (Athens, Greece)
  \emph{(\bibinfo{series}{ASSETS '22})}. \bibinfo{publisher}{Association for
  Computing Machinery}, \bibinfo{address}{New York, NY, USA}, Article
  \bibinfo{articleno}{24}, \bibinfo{numpages}{18}~pages.
\newblock
\showISBNx{9781450392587}
\urldef\tempurl%
\url{https://doi.org/10.1145/3517428.3544819}
\showDOI{\tempurl}


\bibitem[Guidotti et~al\mbox{.}(2018)]%
        {Guidotti2018XAISurvey}
\bibfield{author}{\bibinfo{person}{Riccardo Guidotti}, \bibinfo{person}{Anna
  Monreale}, \bibinfo{person}{Salvatore Ruggieri}, \bibinfo{person}{Franco
  Turini}, \bibinfo{person}{Fosca Giannotti}, {and} \bibinfo{person}{Dino
  Pedreschi}.} \bibinfo{year}{2018}\natexlab{}.
\newblock \showarticletitle{A Survey of Methods for Explaining Black Box
  Models}.
\newblock \bibinfo{journal}{\emph{ACM Comput. Surv.}} \bibinfo{volume}{51},
  \bibinfo{number}{5}, Article \bibinfo{articleno}{93} (\bibinfo{date}{aug}
  \bibinfo{year}{2018}), \bibinfo{numpages}{42}~pages.
\newblock
\showISSN{0360-0300}
\urldef\tempurl%
\url{https://doi.org/10.1145/3236009}
\showDOI{\tempurl}


\bibitem[Ippolito et~al\mbox{.}(2022)]%
        {Ippolito2022CreativeWW}
\bibfield{author}{\bibinfo{person}{Daphne Ippolito}, \bibinfo{person}{Ann
  Yuan}, \bibinfo{person}{Andy Coenen}, {and} \bibinfo{person}{Sehmon Burnam}.}
  \bibinfo{year}{2022}\natexlab{}.
\newblock \showarticletitle{Creative Writing with an AI-Powered Writing
  Assistant: Perspectives from Professional Writers}.
\newblock \bibinfo{journal}{\emph{ArXiv}}  \bibinfo{volume}{abs/2211.05030}
  (\bibinfo{year}{2022}).
\newblock


\bibitem[Jiang et~al\mbox{.}(2022)]%
        {Jiang2022PromptMaker}
\bibfield{author}{\bibinfo{person}{Ellen Jiang}, \bibinfo{person}{Kristen
  Olson}, \bibinfo{person}{Edwin Toh}, \bibinfo{person}{Alejandra Molina},
  \bibinfo{person}{Aaron Donsbach}, \bibinfo{person}{Michael Terry}, {and}
  \bibinfo{person}{Carrie~J Cai}.} \bibinfo{year}{2022}\natexlab{}.
\newblock \showarticletitle{PromptMaker: Prompt-Based Prototyping with Large
  Language Models}. In \bibinfo{booktitle}{\emph{Extended Abstracts of the 2022
  CHI Conference on Human Factors in Computing Systems}} (New Orleans, LA, USA)
  \emph{(\bibinfo{series}{CHI EA '22})}. \bibinfo{publisher}{Association for
  Computing Machinery}, \bibinfo{address}{New York, NY, USA}, Article
  \bibinfo{articleno}{35}, \bibinfo{numpages}{8}~pages.
\newblock
\showISBNx{9781450391566}
\urldef\tempurl%
\url{https://doi.org/10.1145/3491101.3503564}
\showDOI{\tempurl}


\bibitem[Lee et~al\mbox{.}(2022)]%
        {Lee2022Evaluating}
\bibfield{author}{\bibinfo{person}{Mina Lee}, \bibinfo{person}{Megha
  Srivastava}, \bibinfo{person}{Amelia Hardy}, \bibinfo{person}{John
  Thickstun}, \bibinfo{person}{Esin Durmus}, \bibinfo{person}{Ashwin
  Paranjape}, \bibinfo{person}{Ines Gerard-Ursin}, \bibinfo{person}{Xiang~Lisa
  Li}, \bibinfo{person}{Faisal Ladhak}, \bibinfo{person}{Frieda Rong},
  \bibinfo{person}{Rose~E. Wang}, \bibinfo{person}{Minae Kwon},
  \bibinfo{person}{Joon~Sung Park}, \bibinfo{person}{Hancheng Cao},
  \bibinfo{person}{Tony Lee}, \bibinfo{person}{Rishi Bommasani},
  \bibinfo{person}{Michael Bernstein}, {and} \bibinfo{person}{Percy Liang}.}
  \bibinfo{year}{2022}\natexlab{}.
\newblock \bibinfo{title}{Evaluating Human-Language Model Interaction}.
\newblock
\newblock
\urldef\tempurl%
\url{https://doi.org/10.48550/ARXIV.2212.09746}
\showDOI{\tempurl}


\bibitem[Liu et~al\mbox{.}(2023)]%
        {Liu2023PPP}
\bibfield{author}{\bibinfo{person}{Pengfei Liu}, \bibinfo{person}{Weizhe Yuan},
  \bibinfo{person}{Jinlan Fu}, \bibinfo{person}{Zhengbao Jiang},
  \bibinfo{person}{Hiroaki Hayashi}, {and} \bibinfo{person}{Graham Neubig}.}
  \bibinfo{year}{2023}\natexlab{}.
\newblock \showarticletitle{Pre-Train, Prompt, and Predict: A Systematic Survey
  of Prompting Methods in Natural Language Processing}.
\newblock \bibinfo{journal}{\emph{ACM Comput. Surv.}} \bibinfo{volume}{55},
  \bibinfo{number}{9}, Article \bibinfo{articleno}{195} (\bibinfo{date}{jan}
  \bibinfo{year}{2023}), \bibinfo{numpages}{35}~pages.
\newblock
\showISSN{0360-0300}
\urldef\tempurl%
\url{https://doi.org/10.1145/3560815}
\showDOI{\tempurl}


\bibitem[Markov et~al\mbox{.}(2023)]%
        {Markov2023Holistic}
\bibfield{author}{\bibinfo{person}{Todor Markov}, \bibinfo{person}{Chong
  Zhang}, \bibinfo{person}{Sandhini Agarwal}, \bibinfo{person}{Tyna Eloundou},
  \bibinfo{person}{Teddy Lee}, \bibinfo{person}{Steven Adler},
  \bibinfo{person}{Angela Jiang}, {and} \bibinfo{person}{Lilian Weng}.}
  \bibinfo{year}{2023}\natexlab{}.
\newblock \bibinfo{title}{A Holistic Approach to Undesired Content Detection in
  the Real World}.
\newblock
\newblock
\showeprint[arxiv]{2208.03274}~[cs.CL]


\bibitem[Norman(2002)]%
        {Norman2002DOET}
\bibfield{author}{\bibinfo{person}{Donald~A. Norman}.}
  \bibinfo{year}{2002}\natexlab{}.
\newblock \bibinfo{booktitle}{\emph{The Design of Everyday Things}}.
\newblock \bibinfo{publisher}{Basic Books, Inc.}, \bibinfo{address}{USA}.
\newblock
\showISBNx{9780465067107}


\bibitem[Weidinger et~al\mbox{.}(2022)]%
        {Weidinger2022Taxonomy}
\bibfield{author}{\bibinfo{person}{Laura Weidinger}, \bibinfo{person}{Jonathan
  Uesato}, \bibinfo{person}{Maribeth Rauh}, \bibinfo{person}{Conor Griffin},
  \bibinfo{person}{Po-Sen Huang}, \bibinfo{person}{John Mellor},
  \bibinfo{person}{Amelia Glaese}, \bibinfo{person}{Myra Cheng},
  \bibinfo{person}{Borja Balle}, \bibinfo{person}{Atoosa Kasirzadeh},
  \bibinfo{person}{Courtney Biles}, \bibinfo{person}{Sasha Brown},
  \bibinfo{person}{Zac Kenton}, \bibinfo{person}{Will Hawkins},
  \bibinfo{person}{Tom Stepleton}, \bibinfo{person}{Abeba Birhane},
  \bibinfo{person}{Lisa~Anne Hendricks}, \bibinfo{person}{Laura Rimell},
  \bibinfo{person}{William Isaac}, \bibinfo{person}{Julia Haas},
  \bibinfo{person}{Sean Legassick}, \bibinfo{person}{Geoffrey Irving}, {and}
  \bibinfo{person}{Iason Gabriel}.} \bibinfo{year}{2022}\natexlab{}.
\newblock \showarticletitle{Taxonomy of Risks Posed by Language Models}. In
  \bibinfo{booktitle}{\emph{2022 ACM Conference on Fairness, Accountability,
  and Transparency}} (Seoul, Republic of Korea) \emph{(\bibinfo{series}{FAccT
  '22})}. \bibinfo{publisher}{Association for Computing Machinery},
  \bibinfo{address}{New York, NY, USA}, \bibinfo{pages}{214–229}.
\newblock
\showISBNx{9781450393522}
\urldef\tempurl%
\url{https://doi.org/10.1145/3531146.3533088}
\showDOI{\tempurl}


\bibitem[Wu et~al\mbox{.}(2022a)]%
        {Wu2022PromptChainer}
\bibfield{author}{\bibinfo{person}{Tongshuang Wu}, \bibinfo{person}{Ellen
  Jiang}, \bibinfo{person}{Aaron Donsbach}, \bibinfo{person}{Jeff Gray},
  \bibinfo{person}{Alejandra Molina}, \bibinfo{person}{Michael Terry}, {and}
  \bibinfo{person}{Carrie~J Cai}.} \bibinfo{year}{2022}\natexlab{a}.
\newblock \showarticletitle{PromptChainer: Chaining Large Language Model
  Prompts through Visual Programming}. In \bibinfo{booktitle}{\emph{Extended
  Abstracts of the 2022 CHI Conference on Human Factors in Computing Systems}}
  (New Orleans, LA, USA) \emph{(\bibinfo{series}{CHI EA '22})}.
  \bibinfo{publisher}{Association for Computing Machinery},
  \bibinfo{address}{New York, NY, USA}, Article \bibinfo{articleno}{359},
  \bibinfo{numpages}{10}~pages.
\newblock
\showISBNx{9781450391566}
\urldef\tempurl%
\url{https://doi.org/10.1145/3491101.3519729}
\showDOI{\tempurl}


\bibitem[Wu et~al\mbox{.}(2022b)]%
        {Wu2022AIChains}
\bibfield{author}{\bibinfo{person}{Tongshuang Wu}, \bibinfo{person}{Michael
  Terry}, {and} \bibinfo{person}{Carrie~Jun Cai}.}
  \bibinfo{year}{2022}\natexlab{b}.
\newblock \showarticletitle{AI Chains: Transparent and Controllable Human-AI
  Interaction by Chaining Large Language Model Prompts}. In
  \bibinfo{booktitle}{\emph{Proceedings of the 2022 CHI Conference on Human
  Factors in Computing Systems}} (New Orleans, LA, USA)
  \emph{(\bibinfo{series}{CHI '22})}. \bibinfo{publisher}{Association for
  Computing Machinery}, \bibinfo{address}{New York, NY, USA}, Article
  \bibinfo{articleno}{385}, \bibinfo{numpages}{22}~pages.
\newblock
\showISBNx{9781450391573}
\urldef\tempurl%
\url{https://doi.org/10.1145/3491102.3517582}
\showDOI{\tempurl}


\bibitem[Yang et~al\mbox{.}(2022)]%
        {Yang2022ActiveWriter}
\bibfield{author}{\bibinfo{person}{Daijin Yang}, \bibinfo{person}{Yanpeng
  Zhou}, \bibinfo{person}{Zhiyuan Zhang}, \bibinfo{person}{Toby Jia-Jun Li},
  {and} \bibinfo{person}{Ray LC}.} \bibinfo{year}{2022}\natexlab{}.
\newblock \showarticletitle{AI as an Active Writer: Interaction strategies with
  generated text in human-AI collaborative fiction writing}. In
  \bibinfo{booktitle}{\emph{Joint Proceedings of the ACM IUI Workshops}},
  Vol.~\bibinfo{volume}{10}.
\newblock


\bibitem[Yuan et~al\mbox{.}(2022)]%
        {Coenen2021Wordcraft}
\bibfield{author}{\bibinfo{person}{Ann Yuan}, \bibinfo{person}{Andy Coenen},
  \bibinfo{person}{Emily Reif}, {and} \bibinfo{person}{Daphne Ippolito}.}
  \bibinfo{year}{2022}\natexlab{}.
\newblock \showarticletitle{Wordcraft: Story Writing With Large Language
  Models}. In \bibinfo{booktitle}{\emph{27th International Conference on
  Intelligent User Interfaces}} (Helsinki, Finland) \emph{(\bibinfo{series}{IUI
  '22})}. \bibinfo{publisher}{Association for Computing Machinery},
  \bibinfo{address}{New York, NY, USA}, \bibinfo{pages}{841–852}.
\newblock
\showISBNx{9781450391443}
\urldef\tempurl%
\url{https://doi.org/10.1145/3490099.3511105}
\showDOI{\tempurl}


\end{thebibliography}
\appendix
\end{document}